\documentclass[aps,prb,reprint,superscriptaddress,twocolumn,showpacs]{revtex4-1}
\setcitestyle{numbers,square}

\usepackage[T1]{fontenc}
\usepackage[utf8]{inputenc}
\usepackage{textcomp}
\usepackage{color}
\usepackage{comment}
\usepackage{braket}
\usepackage{amsmath}
\usepackage{array,mathtools,amssymb,booktabs}
\usepackage{multirow}
\usepackage{verbatim}

\usepackage{gensymb}

\usepackage{xcolor}

\usepackage{hyperref}
\hypersetup{
  colorlinks=true,
  linkcolor	=black,
  citecolor	=blue,
  urlcolor	=blue
}

\newcommand{\muB}{\mu_\mathrm{B}}

\newcommand{\EF}{E_\mathrm{F}}

\newcommand{\JH}{J_\mathrm{H}}
\newcommand{\diff}{\mathrm{d}}

\newcommand{\AAmath}{\text{\AA}}
\newcommand{\sqsq}{ \sqrt{2} \times \sqrt{2} }
\newcommand{\lainaus}[1]{``#1''}

\usepackage{xcolor}
\definecolor{green2}{RGB}{50,160,50}

\begin{document}

\title{\emph{Ab initio} description of the Bi$_2$Sr$_2$CaCu$_2$O$_{8+\delta}$ electronic structure}

\author{Johannes Nokelainen}
\email[]{johannes.nokelainen@lut.fi}
\affiliation{Department of Physics, School of Engineering Science, LUT University, FI-53850 Lappeenranta, Finland}
\author{Christopher Lane}
\email{laneca@lanl.gov}
\affiliation{Theoretical Division, Los Alamos National Laboratory, Los Alamos, New Mexico 87545, USA}
\affiliation{Center for Integrated Nanotechnologies, Los Alamos National Laboratory, Los Alamos, New Mexico 87545, USA}
\author{Robert S. Markiewicz}
\affiliation{Physics Department, Northeastern University, Boston, Massachusetts 02115, USA}
\author{Bernardo Barbiellini}
\affiliation{Department of Physics, School of Engineering Science, LUT University, FI-53850 Lappeenranta, Finland}
\affiliation{Physics Department, Northeastern University, Boston, Massachusetts 02115, USA}
\author{Aki Pulkkinen}
\affiliation{D\'{e}partement de Physique and Fribourg Center for Nanomaterials, Universit\'{e} de Fribourg, CH-1700 Fribourg, Switzerland}
\affiliation{Department of Physics, School of Engineering Science, LUT University, FI-53850 Lappeenranta, Finland}
\author{Bahadur Singh}
\affiliation{Physics Department, Northeastern University, Boston, Massachusetts 02115, USA}
\author{Jianwei Sun}
\affiliation{Department of Physics and Engineering Physics, Tulane University, Louisiana 70118 New Orleans, USA}
\author{Katariina Pussi}
\affiliation{Department of Physics, School of Engineering Science, LUT University, FI-53850 Lappeenranta, Finland}
\author{Arun Bansil}
\affiliation{Physics Department, Northeastern University, Boston, Massachusetts 02115, USA}

\date{\today}

\begin{abstract} 
Bi-based cuprate superconductors are important materials 
for both fundamental research and applications. 
As in other cuprates, 
the superconducting phase in the Bi compounds 
lies close to an antiferromagnetic phase. 
Our density functional theory calculations based on the 
strongly-constrained-and-appropriately-normed (SCAN) exchange correlation functional 
in Bi$_2$Sr$_2$CaCu$_2$O$_{8+\delta}$ reveal the persistence 
of magnetic moments on the copper ions for oxygen concentrations 
ranging from the pristine phase to the optimally hole-doped compound. 
We also find the existence of ferrimagnetic solutions in the heavily doped compounds, 
which are expected to suppress superconductivity. 
\end{abstract}

\maketitle

\section{Introduction \label{sec:intro}}
In 1986 superconductivity above 30\,K was reported in La$_2$CuO$_4$ by Bednorz and M\"uller~\cite{bednorz1986possible}, 
initiating an intense effort to understand its microscopic origin and 
gain insight into driving the $T_\mathrm{c}$ above the room temperature. 
The anomalous nature of the cuprate superconductivity is believed to originate 
from the quasi-two-dimensional CuO$_2$ planes wherein a strong long-range antiferromagnetic (AFM) 
order is found in the parent half-filled compound~\cite{Tranquada_2014_superconductivity_AFM_neutron}. 
On doping, 
the AFM order quickly disappears and gives way to a superconducting dome. 
From this intimate connection between antiferromagnetism and superconductivity, 
the view that spin-fluctuations play a central role in determining the physical properties of the cuprates has been gaining increasing support~\cite{scalapino1986d,scalapino1995case,2012_Scalapino_RevModPhys_pairing_interaction_spin_fluctuation_superconductor}.  
However, 
there is still no universally accepted explanation for high-temperature superconductivity. 

Crucial to understanding the origin of superconductivity in the cuprates is the process by which doped hole carriers are introduced into the CuO$_2$ planes. 
In simplified low-energy effective models, 
such as the one-band Hubbard model, 
only the Cu-$d$ and O-$p$ states are assumed to dominate. 
This view of the cuprates has been successful in describing the robust broken symmetry phases seen in experiments 
but it does not account for the diversity of transition temperatures at optimal doping. 
For example, 
the highest $T_\mathrm{c}$ obtained in La$_{2-x}$Sr$_x$CuO$_4$ is $40$\,K, 
whereas in the single-layer Hg cuprate, HgBa$_2$CuO$_4$, 
the optimal $T_\mathrm{c}$ is almost $100$\,K, 
more than twice that of La$_{2-x}$Sr$_x$CuO$_4$. 
These variations have been accounted for by modifying 
the local crystal-field splittings in the CuO$_6$ octahedra~\cite{pavarini2001band}, 
which in turn alter fine features of the Fermi surface~\cite{sakakibara2010two}. 
However, 
these models ignore impurity and structural effects derived from real dopants. 
Moreover, 
such models do not account for interlayer coupling effects 
between the CuO$_2$ planes and the charge-reservoir layers. 
Therefore, 
the doping process must be theoretically modeled in a holistic manner by treating the CuO$_2$ plane, 
the surrounding layers, 
and the dopants on the same footing. 

The bismuth-based cuprates Bi$_2$Sr$_2$Ca$_{n-1}$Cu$_n$O$_{2n+4+\delta}$ 
(BSCCO)~\cite{1988_Subramarian_Science_Bi2212_discovery,
1988_Tarascon_BSCCO_110K-Tc,1988_Tarascon_BSCCO_preparation_structure_properties,
2017_Sterpetti_Bi2212_phase_diagram,
2019_FrankZhao_PRL_monolayerBi2212_sign-reversing_hall}
are among the most extensively investigated superconductors. 
Notably, 
the weak van der Waals-like coupling between the layers, 
facilitates cleaving 
and makes BSCCO amenable to accurate 
angle resolved photoemission spectroscopy (ARPES)~\cite{
1994_Aebi_PRL_Bi2212_fermi_surface,
1997_Saini_PRL_ARPES_Bi2212_pseudogap_shadow_bands,
2001_Ding_PRL_ARPES_Coherent_Quasiparticle_Weight,
2002_Lang_Nature_Bi2212_ARPES_granular_structure,
2003_Damascelli_cuprate_megareview,
2015_Lanzara_ARPES_unoccupied_states,
2018_Lanzara_spin-momentum-locking_Bi2212,
2019_Zhou_ARPES_Bi2201_Lifshitz_transition,
2019_Zhou_Bi2212_Arpes_fermi_surface_SC_gap} and 
scanning-tunneling microscopy/spectroscopy (STM/STS)
~\cite{1994_Bernardo_cuprate_gap_theoretical,
2003_McElroy_Nature_Bi2212_ARPES,
2006_Fang_Bi2212_gap-inhomogeneity-induced_electronic_states,
2007_RevModPhys_Renner_cuprate_STM,
2009_Dudy_BSCCO_superstructure_charge_order,
2012_Jouko_PRB_STM_VHS,
2015_Mistark+Bob+Arun_Nanoscale_phase_separation_Bi2201_Ca2CuO2Cl2,
2018_Kuo_Bi2212_ARPES_soft_xray_standing-wave_photoemission,
2018_Li_Bi2212_ARPES_coherent_organization_of_electronic_correlations,
2019_Yu_Nature_monolayerBi2212} studies. 
The two-layer compound ($n=2$) is composed of a rock-salt SrO-BiO$_\delta$-SrO 
charge reservoir layer stacked with two CuO$_2$-Ca-CuO$_2$ layers. 
Unlike the mercury- or yttrium-based cuprates, 
the oxygen impurities in BSCCO can occupy at least three distinct sites. 
These sites have been extensively studied with STM, 
and the findings have been compared to various 
models~\cite{2012_Zeljkovic_Oint_imaging,2014_Zeljkovic_bisco_Oint_periodicity,
2019_Song_2212_Oint+Superstructure_STM_DFT}. 

Initial theoretical studies of cuprates using the density functional theory (DFT) 
missed important Coulomb correlation effects~\cite{1989_Pickett_review_cuprate_electronic_structure}. 
In BSCCO, 
the local-spin-density-approximation (LSDA) fails to produce the copper magnetic 
moments~\cite{
    1988_Massidda_Bi2212_bandstructure,
    1988_Mattheiss_Bi2212_LSDA,
    1988_Hybertsen_Bi2212_bands_LSDA,
    1991_Chan_PRL_Bi2212_positron_annihilation_DFT,
    2006_LinHsin_bisco_BiO_bands}. 
The generalized-gradient-approximation (GGA) produces only marginal corrections 
to the LSDA~\cite{
    2006_He_2212_Oint_DFT,
    2008_He_2212_supermodulation,
    2010_Foyevtsova_2212_doping_DFT,
    2017_Camargo-Martinez_2223_BiO_bands_Pb_doping_DFT,
    2019_Song_2212_Oint+Superstructure_STM_DFT}. 
Jarlborg has suggested applying higher-order density gradient corrections 
to cuprates~\cite{2009_Jarlborg_cuprate_beyond_GGA}. 
Additional studies beyond the GGA using schemes such as 
DFT$+U$~\cite{1991_Anisimov_mott_insulator_DFT+U_theory} 
and DFT$+$DMFT~\cite{2017_Weber_review_cuprate_Tc_STM} 
have been performed to stabilize the AFM ground state. 
However, 
these methods require the use of external parameters such as the Hubbard $U$, 
which limits the predictive power of the theory.

Recent progress 
on advanced DFT schemes offers new pathways for describing 
the electronic structure of correlated materials from first principles. 
In particular, 
the strongly-constrained-and-appropriately-normed (SCAN) 
meta-GGA exchange-correlation functional~\cite{2015_Sun_SCAN}, 
which obeys all known constraints applicable to meta-GGA, 
has been shown to accurately predict many of the key properties of pristine and doped 
La$_2$CuO$_4$~\cite{2018_Chris_La2CuO4_SCAN,2018_Furness_SCAN_cuprate,
2020_Kanun_LSCO_functional_comparisons_arxiv} 
and YBa$_2$Cu$_3$O$_6$~\cite{
2020_Yubo_SCAN_stripe_cuprates}. 
In La$_2$CuO$_4$,
SCAN correctly captures the magnetic moment in magnitude and orientation, 
the magnetic exchange coupling parameter, 
and the magnetic form factor along with the electronic band gap, 
all in accord with the corresponding experimental results.
Ref.~\cite{2020_Kanun_LSCO_functional_comparisons_arxiv} 
compares SCAN with other meta-GGA and hybrid functionals in cuprates and shows that 
SCAN gives the best overall agreement with experiments.
In a SCAN-based study, 
Ref.~\cite{2020_Yubo_SCAN_stripe_cuprates} 
identifies a landscape of 26 competing uniform and stripe phases in near-optimally doped YBa$_2$Cu$_3$O$_7$. 
In Ref.~\cite{2020_Yubo_SCAN_stripe_cuprates}, 
the charge, 
spin and lattice degrees of freedom are treated on an equal footing in a fully self-consistent manner 
to show how stable stripe phases can be obtained without invoking any free parameters. 
These results indicate that SCAN correctly captures many key features 
of the electronic and magnetic structures of the cuprates 
and it thus provides a next-generation baseline for incorporating 
the missing many-body effects such as the quasiparticle lifetimes 
and waterfall-features~\cite{2014_Das+Bob+Arun_cuprate_coupling_model_review}. 
The applicability of SCAN to transition-metal oxides, 
semiconductors, and atomically thin films beyond graphene has been demonstrated 
in Refs.~\cite{varignon2019mott,zhang2019symmetry,2019_Bernardo_Hasnain_SCAN_LixMn2O4,
2016_Sun_SCAN_benchmarking,2016_Car_SCAN_Jacobs_ladder,2017_Buda_SCAN_thin_film,
2018_Isaacs_SCAN_performance_solids,2018_Yubo_NpjComputMater_SCAN_chemical_accuracy,
2020_Chris_Sr2IrO4_SCAN}. 
We note that SCAN also contains overcorrections to the GGA 
in dealing with itinerant ferromagnetism~\cite{
    2018_Singh_SCAN_TM_magnetism,
    2020_Singh_SCAN_magnetism_shortcomings}, 
but the underlying deficiencies responsible for these issues with SCAN 
have been identified and possible fixes have been proposed~\cite{
    2019_Furness_SCAN_iso-orbital_indicator,
    2019_Trickey_SCAN_TM_overmagnetization}. 
There is no evidence that these issues with SCAN persist outside of the ferromagnets, 
since SCAN clearly improves GGA 
in the case of antiferromagnetic $\alpha$-Mn~\cite{2020_oma_Mn}. 

In this article, 
we utilize the SCAN functional to explore the electronic, 
structural and magnetic properties of Bi$_2$Sr$_2$CaCu$_2$O$_8$ (Bi2212) 
on a first-principles basis. 
A realistic description of the phase diagram of BSCCO requires also an accurate treatment 
of the self-doping by the BiO layers and a precise description of the oxygen interstitials, 
which can occupy different sites. 
We will show that a robust copper magnetic moment persists even when 
a substantial amount of oxygen is added to the material, 
which is in agreement with recent resonant inelastic x-ray spectroscopy (RIXS) experiments~\cite{
2014_Grioni_NatCommun_Bi2212_RIXS_doping_anisotropic_softening_of_magnetic_excitations,
2016_Ghiringhelli_RIXS_Bi2201_doping_dependence,
2017_Ghiringhelli_NatPhys_2201_J+others,
2018_Ghiringelli_Bi2212_spin_excitations}. 
The appearance of the Cu magnetic moment in SCAN 
continues to capture other good trends 
seen in the LDA and GGA computations~\cite{2006_LinHsin_bisco_BiO_bands}. 
Finally, 
we find that SCAN predicts ferrimagnetic solutions in overdoped BSCCO 
in agreement with recent experiments 
by Kurashima {\em et al.}~\cite{2018_Kurashima_PRL_BSCCO_FM_fluctuations}. 

This paper is organized as follows. 
Sec.~\ref{sec-methods} discusses the methodology, 
where Sec.~\ref{sec:methods:dft} describes the computational details 
and Sec.~\ref{sec-methods-structure} considers the structural models for BSCCO. 
Sec.~\ref{sec-results} presents the results of this study. 
Here, 
Sec.~\ref{sec-copper} focuses on pristine Bi2212
while Sec.~\ref{sec_typeB_happi} and Sec.~\ref{sec-doping} present the results for oxygen-doped BSCCO 
with O impurities located at various positions in the lattice. 
Sec.~\ref{sec:conclusions} summarizes our conclusions 
and comments on future implications of our work. 

\section{Methodology}\label{sec-methods}

\subsection{Computational details}\label{sec:methods:dft}

{\it Ab initio} calculations were carried out using the projector-augmented-wave 
method~\cite{Blochl1994_PAW,Kresse1999_PAW} as implemented 
in the Vienna ab initio simulation package~\cite{Kresse1996_VASP_CMS,Kresse1996_VASP_PRB}. 
The Kohn-Sham orbitals~\cite{Kohn1998_nobel_lecture_DFT} were expanded in a plane-wave basis set 
with an energy cutoff of 550\,eV. 
The exchange-correlation energy is treated within the SCAN meta-GGA scheme~\cite{2015_Sun_SCAN}. 
Some calculations were also carried out 
within the GGA scheme of Perdew, Burke and Ernzerhof~\cite{Perdew1996_PBE} for reference. 
All sites in the unit cell along with the unit cell dimensions 
were relaxed using a quasi-Newton algorithm to minimize energy 
with an atomic force tolerance of $0.001\,\text{eV}/\text{\r{A}}$. 
A $9\times9\times2$ ($4\times4\times1$) $k$-mesh 
was used to sample the Brillouin zone of the bulk (slab) crystal structure 
and a denser $15\times15\times3$ k-mesh was employed to calculate the density of states (DOS). 
A total energy tolerance of 10$^{-5}$\,eV was used to determine the self-consistent charge density. 
The band structure was unfolded~\cite{
    2013Allen_unfolding_electrons_phonons_slabs,
    2014Medeiros_unfolding1_graphene} 
from the supercell into the primitive cell Brillouin zone 
using the PyProcar~\cite{pyprocar} code. 
Various site-resolved projections were analyzed 
with the pymatgen~\cite{pymatgen} software package.

\subsection{Structural model of BSCCO}\label{sec-methods-structure}

An important characteristic of the cuprates is the presence of an 
intrinsic lattice mismatch between the various 
layers~\cite{2005_Dagotto_Science_Complexity_in_Strongly_Correlated_Electronic_Systems}. 
In BSCCO, 
the substantial tensile stress in the BiO layers leads to an incommensurate 
superlattice modulation~\cite{2011_Poccia_Bi2212_supermodulation_X-ray} 
in which the CuO$_2$ and BiO layers undergo warping and rippling 
with an approximate period of five unit cells along the $b$~axis. 
The reported effects of this supermodulation 
on the electronic properties have been mixed in that 
ARPES finds no effect on $T_\mathrm{c}$ 
as a function of superstructure period~\cite{2019_Liu_and_Zhou_bisco_superstructure_Pb}, 
whereas STM finds the local doping level to be connected to the periodicity 
of the structural modulations~\cite{2020_Zou_PRL_2223_Effect_of_Supermodulation}. 
A few theoretical studies have been performed within the DFT~\cite{
    2008_He_2212_supermodulation,
    2019_Song_2212_Oint+Superstructure_STM_DFT} 
but clear conclusions have been difficult to obtain due to the 
intrinsic limitations of the LSDA and GGA. 

\begin{figure}[htpb]
\centering
\includegraphics[width=\linewidth]{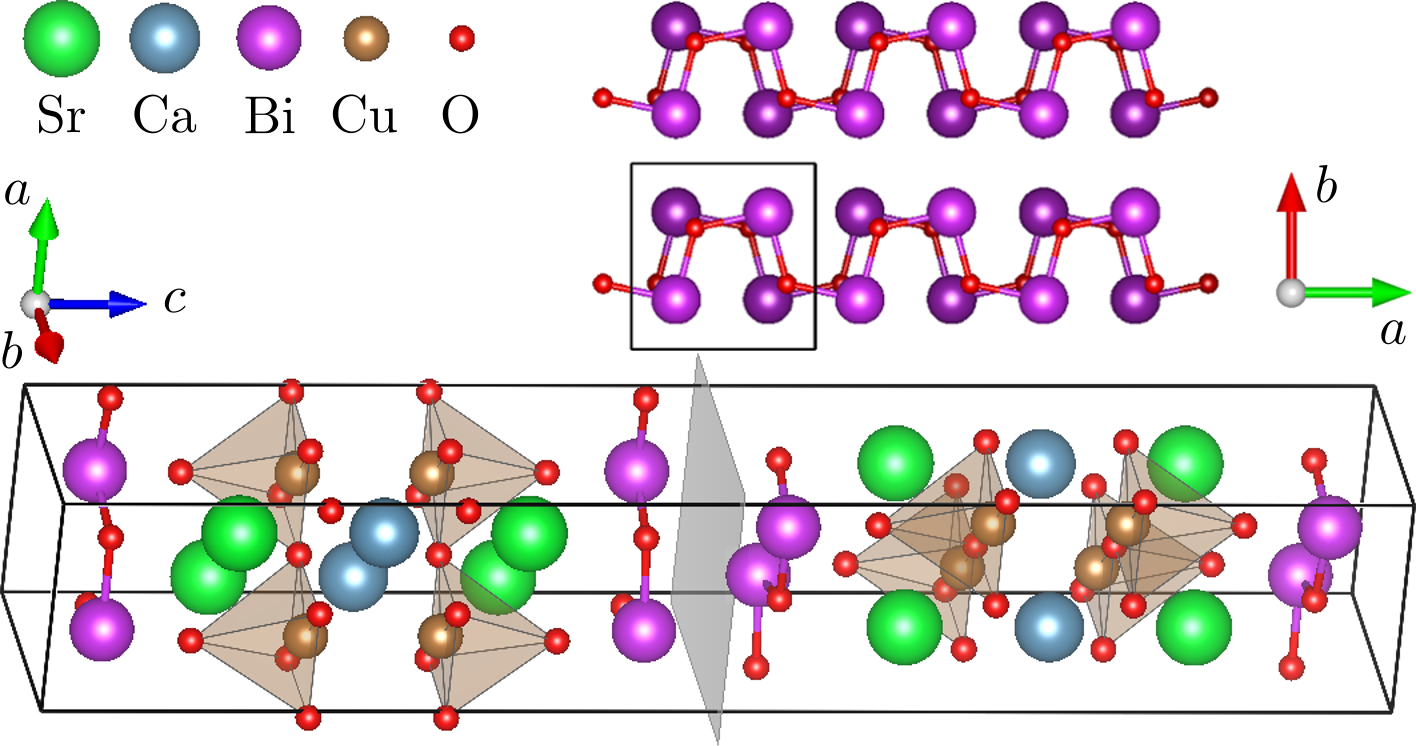}
\caption{\label{fig-structure}
  (Color online). 
  A schematic of the relaxed orthorhombic $\sqsq$ supercell structure 
  of Bi2212 and the zigzag chains of BiO bilayers. 
  This zigzag stacking configuration within the BiO bilayers yielded the lowest energy. 
  The vdW gap in the BiO layer (highlighted with the gray plane) divides the structure into two slabs. 
  Black lines mark the computational unit cell. 
  }
\end{figure}

In this study, 
we neglect the superstructure modulation and focus on the 
electronic and magnetic properties and their evolution with doping. 
In this connection, 
we consider a $\sqsq$ orthorhombic supercell 
to accommodate the $(\pi,\pi)$ AFM order on the copper atomic sites 
(see Fig.~\ref{fig-structure}). 
After relaxing the atomic positions and unit-cell shape, 
we find the $a$, $b$, and $c$ lattice parameters to be 
$5.35\,\AAmath$, $5.42\,\AAmath$, and $31.08\,\AAmath$, respectively, 
admitting a $1.1\,\%$ orthorhombicity in the $ab$~plane. 
These parameters are in good accord with the corresponding experimental results 
[$a=5.399(2)\,\AAmath$, 
$b=5.414(1)\,\AAmath$, 
and $c=30.904(16)\,\AAmath$]~\cite{1988_Subramarian_Science_Bi2212_discovery}. 
Interestingly, 
in comparison to a freestanding BiO bilayer, 
our computations show that the BiO bilayer in BSCCO is under a tensile strain of 9.3\,\% 
due to lattice mismatch~\footnote{We are not aware of experimental lattice parameters 
for a freestanding BiO bilayer in the literature.}.
Consequently,
the Bi and O ions rearrange themselves and exhibit
stronger BiO bonding along the $a$~axis compared to the $b$~axis, 
yielding zig-zag BiO chains or Bi$_2$O$_2$ quadrilaterals~\cite{
    2008_He_2212_supermodulation,
    2011_Fan_BiO_distortion_DFT,
    2012_Zeljkovic_BiO_disorder_bisco}. 
The chain formation appears to be key for stabilizing the orthorhombic ground state. 

In order to delineate effects of doping, 
we doubled the unit cell in the $ab$~plane. 
Since the bulk Bi2212 crystal consists of two formula units 
stacked body-center-wise and separated by a van der Waals (vdW) region 
with very little $k_z$ 
dispersion~\footnote{We studied the $k_z$ dispersion of the electronic structure by looking along the $(0,0,0)-(0,0,\frac{\pi}{2})$ and $(\frac{\pi}{2},\frac{\pi}{2},0)-(\frac{\pi}{2},\frac{\pi}{2},\frac{\pi}{2})$ paths in the  AFM BZ and we found no significant dispersion.},
we followed previous computational 
studies~\cite{2006_He_2212_Oint_DFT,2010_Foyevtsova_2212_doping_DFT,2014_Zeljkovic_bisco_Oint_periodicity}
and considered only one formula unit. 
Using a small vacuum region of $3.8\,\AAmath$ 
to separate the periodic images of these slabs, 
we verified
that the electronic properties of this simplified model 
correspond well to those of the bulk.

\section{RESULTS}\label{sec-results}

\subsection{Electronic structure of pristine Bi2212}\label{sec-copper}

\begin{figure}[htpb]
\includegraphics[width=\linewidth]{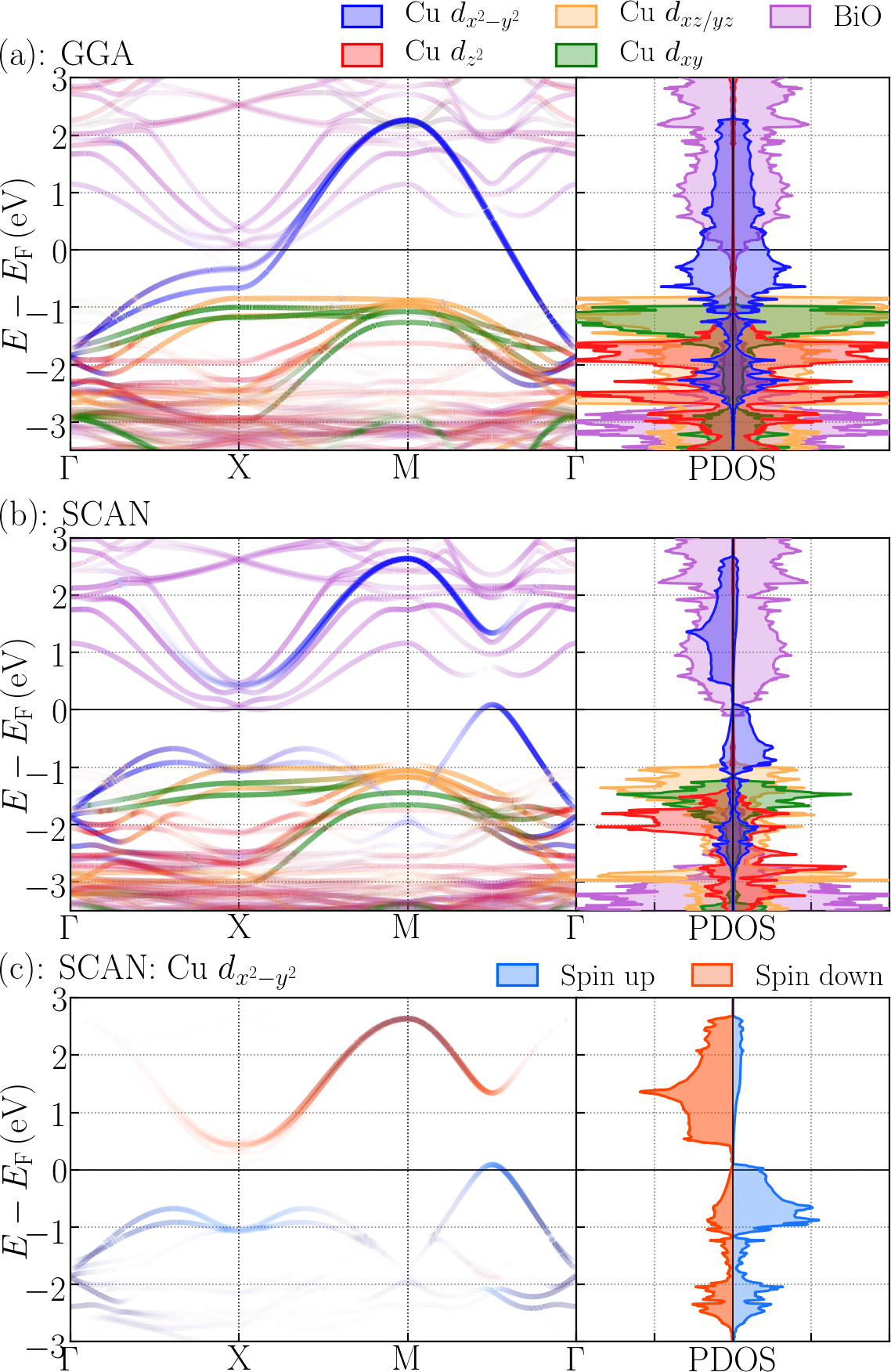}
\caption{\label{fig:PBESCAN}
  [(a) and (b)] 
  Band structure and DOS projected onto Cu $d$ orbitals and BiO layers for GGA and SCAN. 
  For SCAN, 
  the Cu projections refer only to Cu ions with positive magnetic moments 
  in order to highlight their spin polarization in PDOS 
  (the total moment over the unit cell is zero). 
  However, 
  the spin polarization is not shown in the 
  band structure plots for simplicity. 
  (c): SCAN-based Cu $d_{x^2-y^2}$ bands and DOS. 
  Here, 
  spin polarization of the band structure is shown. 
  Band structure has been unfolded~\cite{
    2013Allen_unfolding_electrons_phonons_slabs,
    2014Medeiros_unfolding1_graphene} 
  into the primitive cell 
  from the AFM $\sqrt2 \times \sqrt2$ supercell. 
  $X$ and $M$ symmetry points are given with respect to the Brillouin zone of the primitive cell. 
}
\end{figure}

Fig.~\ref{fig:PBESCAN} 
compares band structures and site-resolved 
partial-densities-of-states (PDOSs) of 
Bi2212 obtained within the GGA and SCAN schemes. 
Consistent with previous \emph{ab initio} 
studies~\cite{2006_LinHsin_bisco_BiO_bands,2010_Foyevtsova_2212_doping_DFT}, 
GGA [Fig.~\ref{fig:PBESCAN}\,(a)] yields a nonmagnetic metal, 
where the spin-degenerate Cu $d_{x^2-y^2}$ bands cross the Fermi level 
with an overall bandwidth of 4.0\,eV. 
In contrast, 
SCAN [Fig.~\ref{fig:PBESCAN}\,(b)] stabilizes the AFM order 
over the copper sublattice and produces 
an indirect gap of 0.33\,eV in the half-filled $d_{x^2-y^2}$-dominated band. 
At the $X$ point, 
the energy gap is 1.47\,eV, 
while at the midpoint between $M$ and $\Gamma$ 
\footnote{In the supercell BZ this point corresponds to the $X$-point, 
and the primitive cell $X$-point corresponds to the supercell $M$ point
},
the gap is 1.24\,eV. 
When the band structure is projected onto the Cu ions 
with positive magnetic moments 
[Fig.~\ref{fig:PBESCAN}\,(c)], 
the spin-polarized nature of the Cu $d_{x^2-y^2}$ bands become visible. 
The valence band (majority spin) is now seen to be partially occupied, 
while the conduction band (minority spin) is unoccupied, 
leading to local Cu magnetic moments of $\pm0.425\,\muB$. 
Around $X$, 
the valence bands exhibit a bilayer splitting of $0.24$\,eV, 
which produces two van Hove singularities 
visible in the PDOS at around $-0.65$\,eV and $-0.88$\,eV. 
These singularities visually appear to be stronger than logarithmic, 
in agreement with Nieminen \emph{et al.}~\cite{2012_Jouko_PRB_STM_VHS}. 
The Cu $d_{z^2}$ and $t_{2g}$ orbitals are spin-split due to the Hund's coupling 
[see Fig.~\ref{fig:PBESCAN}\,(b)]. 
This splitting is substantial for the $d_{z^2}$ orbitals 
but weak for the $t_{2g}$ orbitals; 
we find that the $d_{z^2}$ orbitals contribute mainly 
between $-1.7$\,eV and $-2.2$\,eV in the spin up channel 
and between $-2.7$\,eV and $-3.2$\,eV in the spin down channel. 
In contrast, 
the $t_{2g}$ majority and minority spin states are nearly degenerate, 
with the weight of $d_{xz/yz}$ orbitals concentrated 
between $-1.7\,$eV and $-1.0$\,eV, 
and that of $d_{xy}$ orbitals between $-2.0\,$eV and $-1.2$\,eV. 
Hund's coupling leads to similar orbital splitting behavior in 
La$_2$CuO$_4$, 
but with different ordering of the $d$ orbitals~\cite{2018_Chris_La2CuO4_SCAN}. 
Here, 
the $d_{z^2}$ bands are below the $t_{2g}$ bands, 
whereas in La$_2$CuO$_4$ they are the highest fully occupied bands. 
This difference between Bi2212 and La$_2$CuO$_4$ 
is a consequence of the larger separation 
between the Cu ions and the apical oxygen atoms in Bi2212; 
2.67\,{\AA} in our relaxed structure 
compared to 2.45\,{\AA} in La$_2$CuO$_4$~\cite{2018_Chris_La2CuO4_SCAN}. 

In order to estimate the value of on-site Hubbard potential $U$ 
and the Hund's coupling $\JH$, 
we follow the approach of Lane \emph{et al.}~\cite{2018_Chris_La2CuO4_SCAN}. 
Using the PDOSs $g_{\mu\sigma}$ resolved by orbitals $\mu$ and spin $\sigma$, 
we determine the average spin-splitting $\overline{E}_{\mu\sigma}$ 
of the $d_{x^2-y^2}$ and $d_{z^2}$ levels 
and then $U$ and $\JH$ as follows:
\begin{align}
\label{eq1}
\overline{E}_{\mu\sigma}
&=
\int_W E g_{\mu\sigma}(E) \,\diff E,
\\\label{eq2}
\overline{E}_{d_{x^2-y^2}\uparrow} - \overline{E}_{d_{x^2-y^2}\downarrow}
&=
U (N_\uparrow-N_\downarrow),
\\\label{eq3}
\overline{E}_{\mu\neq d_{x^2-y^2}\uparrow} - \overline{E}_{\mu\neq d_{x^2-y^2}\downarrow}
&=
\JH(N_\uparrow - N_\downarrow),
\end{align}
where $N_\uparrow$ ($N_\downarrow$) is the occupation of the spin-up (down) 
$d_{x^2-y^2}$ orbital and the integration is over the full bandwidth $W$. 
In this way,
$U$ and $\JH(\mu = d_{z^2})$ are found to be $4.7\,$eV and 1.35\,eV, 
respectively. 
These values are very similar to those found for La$_2$CuO$_4$~\cite{2018_Chris_La2CuO4_SCAN}. 
Also, 
this value of $U$ is comparable to that found 
in the 3-band Hubbard models of cuprates, 
but it is substantially larger than the $U$ used in the single-band Hubbard model,
which can be estimated through a constrained random phase approximation 
calculation~\cite{2016_Jang_cRPA_Hg-cuprates}
for Bi2212~\cite{2017_Yang_Coulomb_U_experiment_2212}
and Bi$_2$Sr$_2$CuO$_6$ (Bi2201)~\cite{2017_Ghiringhelli_NatPhys_2201_J+others}. 
This difference is due to the over-simplified 
nature of the single-band model, 
where the band is composed of Cu-$d_{x^2-y^2}$ and O-$p_{x},p_{y}$ characters. 
This band thus essentially represents a CuO$_2$ molecule 
instead of a pure $d$ state, 
so that the $U$ estimated in this way involves partial screening by the O ligands. 

The nearest-neighbor super-exchange coupling parameter $J$ 
is usually estimated by mapping to an effective 
Heisenberg model~\cite{2018_Chris_La2CuO4_SCAN}. 
However, 
this is not possible here because we found that the 
ferromagnetic state in this case converges to zero magnetic moment. 
For this reason, 
we have used $J \approx 4t^2/U - 24t^4/U^3$, 
where $t$ is the nearest-neighbor hopping parameter, 
which can be estimated from the $d_{x^2-y^2}$ bandwidth $B$ 
to be $t = B/8 \approx 500$\,meV. 
We thus estimate $J\approx200$\,meV,
which is in reasonable accord with the corresponding experimental value 
of $\sim$\,148\,meV~\cite{2017_Ghiringhelli_NatPhys_2201_J+others}. 

Unlike the other cuprates such as La$_2$CuO$_4$, 
pristine BSCCO is weakly metallic due to self-doping~\cite{2006_LinHsin_bisco_BiO_bands}: 
both the BiO and Cu $d_{x^2-y^2}$ bands cross $\EF$ 
and lead to a semimetal through the removal of some electrons from the CuO plane. 
This self-doping effect may be the reason that it has been difficult 
to stabilize a large magnetic gap in nominally pristine BSCCO 
without rare-earth substitution~\cite{1994_Moudden_2212_with_rare_earth_neutron_antiferro,
1995_Awana_2212_wit_rare_earth_synthesis_magnetism}. 
We have also carried out computations on Bi2201
(see Supplementary Materials~\cite{
    Supplementary,
    1988_Imai_2201_2223_structure} 
for comparison of Bi2201 with Bi2212). 
Notably, 
the Cu magnetic moment in Bi2201 is found to be $0.395\,\muB$, 
which is $0.030\,\muB$ less than in Bi2212. 
This reflects the effect of stronger 
self-doping in Bi2201 where the Bi/Cu ratio is twice as large as in Bi2212.

\subsection{Doping of Bi2212}\label{sec-doping}

\begin{figure}
\centering
\includegraphics[width=.76\linewidth]{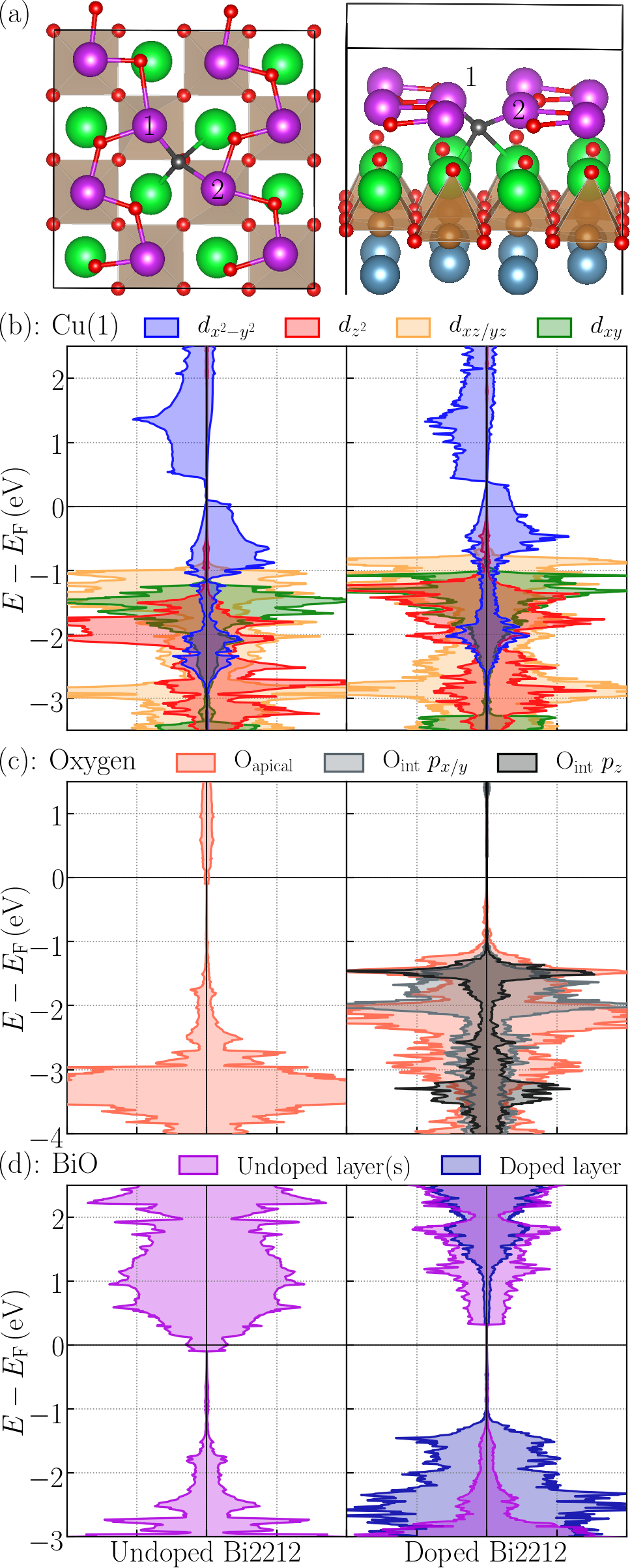}
\caption{\label{fig-happi}
  (a) Top and side views of the relaxed Bi2212 slab 
  with an type A-type O dopant. 
  O$_\text{int}$ is colored black. 
  The Cu$(1)$/Cu$(2)$ and the O$_\text{apical}(1)$/O$_\text{apical}(1)$ atoms 
  are located below the Bi atoms labeled 1/2. 
  (b) $lm$-decomposed PDOS of the Cu$(1)$ ion 
  [closest to O$_\text{int}$, see panel (a)] 
  in pristine (left) and doped (right) cases. 
  (c) $p$-projected PDOS of the O$_\text{apical}(1)$ atom 
  with and without doping. 
  For the doped case (right), 
  the PDOS of the O$_\text{int}$ is also presented. 
  (d) PDOS projected onto the BiO layers.
}
\end{figure}

STM studies of Zeljkovic \emph{et al.}~\cite{2012_Zeljkovic_Oint_imaging,2014_Zeljkovic_bisco_Oint_periodicity} 
show that there are two different types of interstitial oxygen dopants in BSCCO. 
The \lainaus{type B} dopants reside in the BiO layers, 
whereas the \lainaus{type A} oxygens lie close to the apical oxygen atoms 
and the SrO layers and interact directly with the CuO$_2$ planes. 
We have modeled both types of these dopants and found that the type A oxygen 
dopants explain most of the observed hole-type doping. 
The B oxygen dopants are discussed further in Sec.~\ref{sec_typeB_happi} below. 
Our calculations for modeling doping effects
employed a 120-atom $2\sqrt{2}\times2\sqrt{2}$ supercell slab 
(see Sec.~\ref{sec-methods} for details) 
with a type A interstitial oxygen atom O$_\text{int}$. 
This model corresponds to a doping level of $\delta=1/8$ 
(close to optimal doping), 
as illustrated in Fig.~\ref{fig-happi}\,(a). 
In the relaxed structure the O$_\text{int}$ atom is found to reside 
between the SrO and BiO layers 
in agreement with the results of 
He \emph{et al.}~\cite{2006_He_2212_Oint_DFT,2008_He_2212_supermodulation} 
and Foyevtsova \emph{et al.}~\cite{2010_Foyevtsova_2212_doping_DFT} 
as well as with a recent 
scanning-transmission-electron microscopy (STEM) study by Song 
\emph{et al.}~\cite{2019_Song_2212_Oint+Superstructure_STM_DFT}. 

In Fig.~\ref{fig-happi}\,(b), 
we illustrate the effects of dopant on the electronic structure 
by comparing the pristine and doped PDOS on Cu$(1)$ site, 
which is the copper ion closest to the O$_\text{int}$. 
Doping leads to addition of holes resulting into 
a downwards shift of the Fermi level in the magnetic $d_{x^2-y^2}$ band, 
along with the closing of the $d_{x^2-y^2}$ electronic gap. 
The doping leads to a reduction in the average value of the Cu magnetic moment 
($|\overline{M}|=0.347\,\muB$) by $0.078\,\muB$. 
Values of $|M|$ differ significantly between the two CuO$_2$ planes. 
We will refer to the CuO$_2$ planes with/without the dopant 
as \lainaus{doped/undoped} planes. 
On the undoped plane, $|M|=0.363\,\muB$, 
whereas on the doped plane, 
the magnetic moments are on average $0.328\,\muB$ 
with significant variations on Cu sites 
($0.322\,\muB\leq|M|\leq0.339\,\muB$). 
The on-site potential of about 4.8\,eV, 
calculated from the Cu $d_{x^2-y^2}$ PDOS, 
is constant for all the Cu sites 
and remains almost unchanged from the pristine case. 

Note that the O dopant here resides between the apical oxygen atoms 
O$_\text{apical}(1)$ and O$_\text{apical}(2)$  
at distances of $2.61\,\AAmath$ and $2.66\,\AAmath$, 
respectively [see Fig.~\ref{fig-happi}\,(a)]. 
The O$_\text{int}$ interacts with the Cu$(1)$ $d_{z^2}$ orbitals 
primarily through O$_\text{apical}(1)$. 
However, 
as discussed in Sec.~\ref{sec-copper}, 
this interaction is suppressed in BSCCO compared to other cuprates 
due to the larger Cu--O$_\text{apical}$ separation. 
Lack of hybridization in the pristine case can be seen by comparing 
the Cu$(1)$ $d_{z^2}$ and O$_\text{apical}(1)$ PDOSs 
on the left sides of Figs.~\ref{fig-happi}\,(b) and \ref{fig-happi}\,(c). 
In the doped case, 
the coupling between the Cu$(1)$/Cu$(2)$ ions and the 
O$_\text{apical}(1)$/O$_\text{apical}(2)$ atoms 
is significantly enhanced because of 0.18\,{\AA} reduction 
of their separations to 2.49\,{\AA}. 
Consequently, 
the O$_\text{apical}(1)$ states are lifted from below $-3$\,eV 
to the energy interval of $-3$\,eV to $-1$\,eV, 
as shown in Fig.~\ref{fig-happi}\,(c). 
This modified O$_\text{apical}(1)$ PDOS displays strong common features 
with the Cu$(1)$ $d_{z^2}$ states 
as illustrated in the right side of Fig.~\ref{fig-happi}\,(b). 
These results indicate substantial doping-induced 
interactions between these atoms. 
On the Cu$(1)$ ion, 
the effect of these interactions is to lift the 
$d_{z^2}$ orbitals by $\sim0.3\,$eV with respect to the $t_{2g}$ orbitals, 
which can be seen by comparing their average energies 
computed with Eq.~\eqref{eq1}. 
In addition, 
the shape of the Cu$(1)$ $d_{z^2}$ PDOS experiences significant modification. 
However, 
the estimated Hund's splitting ($1.38$\,eV) remains almost unchanged. 
The overall trends described above are also present on the other 
Cu sites in a less pronounced form. 

The right side of Fig.~\ref{fig-happi}\,(c) 
gives insight into the nature of O$_\text{int}$ PDOS. 
By comparing PDOSs of O$_\text{int}$ and O$_\text{apical}(1)$
we see that both $p_{x/y}$ and $p_z$ orbitals of O$_\text{int}$ 
couple with O$_\text{apical}(1)$, 
with $p_z$ coupling around $-1.4\,$eV and $p_{x/y}$ around $-2.0\,$eV. 
The O$_\text{int}$ $p_z$ PDOS is especially relevant for STM experiments 
since the tunneling involves the $p_z$ orbital while the $p_x/p_y$ orbitals 
are orthogonal to the STM tip~\cite{2011_Jouko_STM_tip_orbitalsymmetery_Bi2212}. 
Indeed, 
STM studies by Zeljkovic \emph{et al.}~\cite{
    2012_Zeljkovic_Oint_imaging,
    2014_Zeljkovic_bisco_Oint_periodicity} 
report a peak in the scanning-tunneling spectrum 
at $-1.5\,$eV for the type A interstitial, 
which is close to the aforementioned 
O$_\text{int}$ $p_z$ PDOS peak at $-1.4\,$eV. 

Fig.~\ref{fig-happi}\,(d) 
shows the BiO-layer PDOS with and without the dopants. 
Doping is seen to lift the BiO bands above $\EF$ in accord with the study 
of Lin \emph{et al.}~\cite{2006_LinHsin_bisco_BiO_bands} 
and Bi2223 study of Camargo-Mart\'{i}nez 
\emph{et al.}~\cite{2017_Camargo-Martinez_2223_BiO_bands_Pb_doping_DFT} 
where doping was done with Pb instead of O. 
Note that BiO pockets are removed also from BiO layer 
which does not lie close to the O$_\text{int}$ although 
effects of dopant on this \lainaus{undoped} layer are relatively weak. 
In contrast, 
the dopant induces substantial effects on the electronic states from the 
\lainaus{doped} BiO layer (i.e. the layer close to O$_\text{int}$) 
where the spectral weights associated with the BiO states are lifted upwards 
by more than 1\,eV and the BiO bands now overlap the Cu $d$ bands in energy. 

We also investigated the heavily overdoped regime ($\delta=1/4$) 
by introducing a second type-A dopant 
that was placed in the structure as far as possible from the first dopant. 
Compared to $\delta=1/8$, 
the average value of $|M|$ of the Cu ions in the overdoped case 
is lowered by $0.078\,\muB$ to 0.268\,$\muB$. 
Interestingly, 
the higher doping also leads to the onset of ferrimagnetic order 
with average spin up/down moments on Cu atoms of $0.307\,\muB$\,/\,$-0.229\,\muB$. 
Moreover, 
the oxygen atoms in the CuO$_2$ planes now develop a magnetic moment 
of $+0.010\,\muB$. 
The total magnetization of the unit cell is $0.059\,\muB$ per copper. 
Such magnetization has been predicted to  destroy superconductivity 
in overdoped cuprates~\cite{
    2018_Kurashima_PRL_BSCCO_FM_fluctuations,
    2007_Kopp_PNAS_ferromagnetism_cuprate,
    2008_Bernardo_PRL_ferromagnetism_LSCO}. 

\subsection{Type B oxygen dopants}\label{sec_typeB_happi}

\begin{figure*}[htpb]
\includegraphics[width=\linewidth]{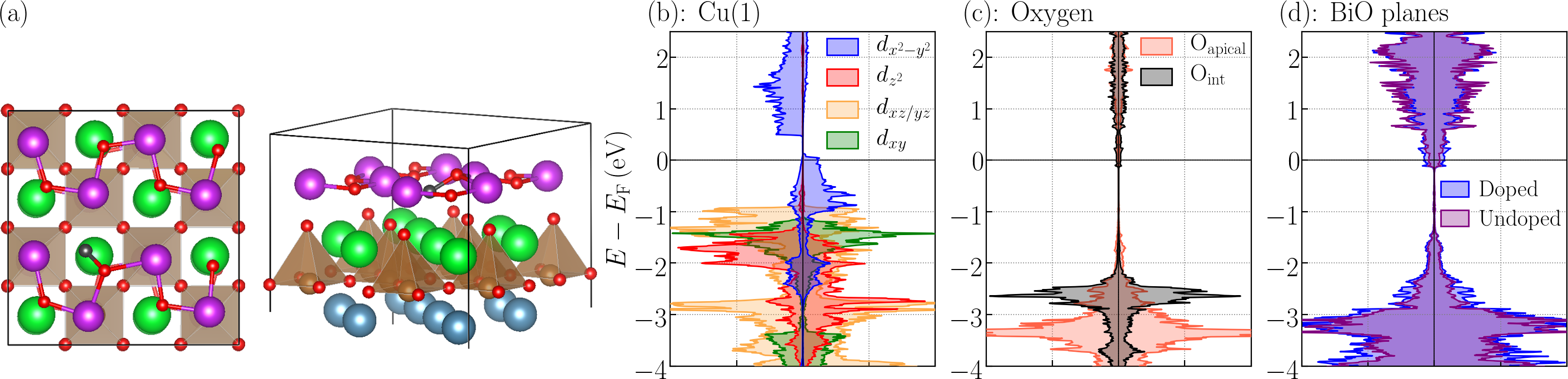}
\caption{\label{fig-happiB}
  (a) Structural model of a type B O-dopant in Bi2212. 
  O$_\text{int}$ is colored black. 
  (b) PDOS of various $d$ orbitals 
  of a copper atom close to the O$_\text{int}$. 
  (c) 
  PDOS of the $p$ orbitals of the O$_\text{int}$ and an apical oxygen atom 
  close to the dopant. 
  (d) PDOS projected onto the BiO layers with and without the dopant.. 
  }
\end{figure*}

Following the experimental results of Zeljkovic 
\emph{et al.}~\cite{
    2012_Zeljkovic_Oint_imaging,
    2014_Zeljkovic_bisco_Oint_periodicity} 
and the computational study of 
He \emph{et al.}~\cite{2006_He_2212_Oint_DFT}, 
we placed the-type B oxygen dopants in the middle of the 
approximately square Bi network (position \#2 of He \emph{et al.}). 
This location is quite close to one of the oxygen atoms in the 
BiO layer and leads to the formation of an oxygen molecule 
as shown in Fig.~\ref{fig-happiB}\,(a). 
We find the bond length of this oxygen dimer to be $1.476\,$\AA, 
which is close to the [O$_2$]$^{2-}$ bond length in BaO$_2$ 
of $1.49\,\AAmath$~\cite{
1958_Sutton_book_interatomic_distances_and_configuration_in_molecules_and_ions}. 
This O$_\text{int}$ stabilizes into a position slightly below the BiO layer, 
while the oxygen which it is attached lies above the BiO layer, 
so that the dimer is tilted by an angle of {33\textdegree} from the $c$~axis. 
The total energy of the type B-doped compound was found to be 
2.27\,eV higher than that of the type A-doped structure. 

In contrast to our results for the type A interstitial O atom, 
we found that the B interstitials produce only little doping, 
with the Cu magnetic moments being decreased 
only by $0.014\,\muB$ to $0.402\,\muB$. 
The Cu $d_{x^2-y^2}$ state remains nearly unchanged, 
as seen from the PDOS in Fig.~\ref{fig-happiB}\,(b), 
and the BiO pocket is not lifted above $\EF$, 
as illustrated in Fig.~\ref{fig-happiB}\,(d). 
In the PDOS of the O$_\text{int}$, 
a clear peak appears at around $-2.6$\,eV. 
This feature is also reflected in the PDOS of the O$_\text{apical}$ 
[see Fig.~\ref{fig-happiB}\,(c)]
and in the PDOS of the Cu $d_{z^2}$ 
[see Fig.~\ref{fig-happiB}\,(b)], 
indicating that some interactions occur 
also between the type B O$_\text{int}$ and the CuO$_2$ plane. 

We also tested the interstitial oxygen position in the van der Waals gap between the BiO layers. 
The energy of this configuration was found to be between 
that of type A and B oxygen atoms. 
To the best of our knowledge,
this impurity position has not been considered in the literature. 
A possible explanation is that these oxygen atoms are very mobile 
and therefore they disappear during the annealing of the material 
or combine with existing oxygens in the BiO layer to become 
type B oxygens. 
Additionally, 
they might be more sensitive to the supermodulation distortions, 
which are not considered in our structural model. 

\section{Summary and Conclusions}\label{sec:conclusions}

We have discussed the electronic structure of BSCCO compounds 
using accurate first-principles computations 
based on the SCAN functional, 
which does not require the introduction of any arbitrary parameters 
(e.g., the Hubbard $U$) 
to describe Coulomb correlation effects. 
As in our previous investigations of various cuprates, 
SCAN is found to greatly improve the description 
of the electronic states in the BSCCO system. 
In particular, 
our results yield accurate lattice geometries, 
copper magnetic moments and band structures that are in better agreement 
with experiments than GGA. 
The copper magnetic moments exhibit an antiferromagnetic coupling 
with and without oxygen dopants in accord with RIXS measurements, 
suggesting that superconductivity could be connected with 
quasiparticles coupled to spin fluctuations\cite{
2012_Scalapino_RevModPhys_pairing_interaction_spin_fluctuation_superconductor}. 
Oxygen dopants are shown to increase the coupling between the apical oxygens 
and the CuO$_2$ layers 
and modify especially the Cu $d_{z^2}$ states. 
We also find the appearance of a doping-induced ferrimagnetic order 
that could be responsible for the suppression of superconductivity 
in the overdoped regime. 
The competition between superconductivity and ferrimagnetism 
hints that further studies of overdoped BSCCO 
could clarify important open questions such as 
the observation of a second dome of higher temperature 
superconductivity in the cuprates~\cite{2019_Chu_2nd_dome_pressure_bisco}. 

\begin{acknowledgments}

It is a pleasure to acknowledge important discussions with 
Giacomo Ghiringhelli, Marco Moretti and Xingjiang Zhou. 
The authors acknowledge CSC-IT Center for Science, Finland, 
for computational resources. 
J.N. is supported by the Finnish Cultural Foundation.
The work at Northeastern University was supported by the US Department of Energy (DOE), 
Office of Science, 
Basic Energy Sciences grant number DE-FG02-07ER46352 (core research), 
and benefited from Northeastern University's Advanced Scientific Computation Center (ASCC), 
the NERSC supercomputing center through DOE grant number DE-AC02-05CH11231, 
and support (testing efficacy of advanced functionals) from the DOE EFRC: Center for Complex Materials from First Principles (CCM) under grant number DE-SC0012575. 
C.L. was supported by the U.S. DOE NNSA under Contract No. 89233218CNA000001 and by the Center for Integrated Nanotechnologies, a DOE BES user facility, in partnership with the LANL Institutional Computing Program for computational resources. Additional support was provided by DOE Office of Basic Energy Sciences Program E3B5. 
B.B. acknowledges support from the COST Action CA16218. 
A.P. acknowledges the Osk. Huttunen Foundation for financial support.
J.S. was supported by the US Department of Energy 
under EPSCoR Grant No.~DE-SC0012432 
with additional support from the Louisiana Board of Regents. 

\end{acknowledgments}

\bibliography{bscco}

\end{document}